# SEAWATER pH AND ANTHROPOGENIC CARBON DIOXIDE

Gerald E. Marsh

Argonne National Laboratory (Ret)
5433 East View Park
Chicago, IL 60615

E-mail: gemarsh@uchicago.edu

**Abstract.** In 2005, the Royal Society published a report titled *Ocean acidification due to increasing atmospheric carbon dioxide*. The report's principal conclusion—that average ocean pH could decrease by 0.5 units by 2100—is demonstrated here to be consistent with a linear extrapolation of very limited data. It is also shown that current understanding of ocean mixing, and of the relationship between pH and atmospheric carbon dioxide concentration, cannot justify such an extrapolation.

**Introduction.**

The 2005 report of the Royal Society maintains that we face a crisis in continued acidification of the oceans due to the emissions of anthropogenic carbon dioxide [1]. They state "Ocean acidification is a powerful reason, in addition to that of climate change, for reducing global $CO_2$ emissions. Action needs to be taken now to reduce global emissions of $CO_2$ to the atmosphere to avoid the risk of irreversible damage to the oceans. We recommend that all possible approaches be considered to prevent $CO_2$ reaching the atmosphere. No option that can make a significant contribution should be dismissed."

Quantitatively, it is further maintained that "If global emissions of $CO_2$ from human activities continue to rise on current trends then the average pH of the oceans could fall by 0.5 units (equivalent to a three fold increase in the concentration of hydrogen ions) by the year 2100. This pH is probably lower than has been experienced for hundreds of millennia and, critically, this rate of change is probably one hundred times greater than at any time over this period."

Zeebe, et al. [2] argue that, in contrast to climate model predictions, "future ocean chemistry projections are largely model-independent on a time scale of a few centuries, mainly because the chemistry of $CO_2$ in seawater is well known and changes in surface ocean carbonate chemistry closely track changes in atmospheric $CO_2$."

These are very strong claims and if true mandate immediate action to protect one of the world's most precious assets, one that is currently experiencing the tragedy of the commons due to over fishing and habitat destruction. There are, however, significant uncertainties in projections of ocean pH even though "the chemistry of $CO_2$ in seawater is well known".

The organization of this paper is as follows: The relationship between pH and $pCO_2$ is discussed in terms of two expressions easily derived from the chemistry of seawater (the derivations are given in the Appendix); this is followed by a description of the limited pH data available from Ocean Station Aloha; a comparison of the actual data with the theoretical models follows. It is then shown that the extrapolation of pH in the Royal Society Report is consistent with a doubtful linear extrapolation of the Ocean Station Aloha or similar data. There follows a discussion of ocean mixing times and the Royal Society assumption that there will be no mixing of surface water with the deep ocean. An introduction to seawater response to increased atmospheric carbon dioxide concentrations is also given in the Appendix.

**The relationship between pH and pCO₂**

In footnote 32, Pearson and Palmer [3] give the following equation for estimating $pCO_2$ as a function of hydrogen ion concentration and total dissolved carbon:

$$pCO_2 = K_H[H_2CO_3] = K_H \left(1 + \frac{K_1}{[H^+]} + \frac{K_1 K_2}{[H^+]^2}\right)^{-1} \Sigma CO_2.$$

(1)

Order of magnitude estimates of the dissociation constants imply that only the second term in the denominator of the r.h.s. is of importance. This results in the relation

$$pCO_2 = \frac{K_H \Sigma CO_2}{K_1 10^{pH}}.$$

(2)

Caldeira and Berner [4] give the expression

$$pCO_2 = \frac{K_H}{[K_1]}[H^+][HCO_3^-] = \frac{K_H}{K_1 K_2}[H^+]^2[CO_3^{2-}],$$

(3)

which implies that

$$pCO_2 = \frac{K_H}{K_1 K_2} \frac{[CO_3^{2-}]}{100^{pH}}.$$

(4)

Pearson and Palmer hold $\Sigma CO_2$ constant so that $pCO2 \propto [H^+]$, while Caldeira and Berner hold $[CO_3^{2-}]$ constant so that $pCO2 \propto [H^+]^2$.

Eqs. (2) and (4) are two simple proportionality relations between $pCO_2$ and pH. If one now chooses the proportionality constants such that each relation gives the assumed pre-industrial pH of 8.25 at the pre-industrial $pCO_2$ value of 280 ppmv and plots the result, one obtains Fig. 1. Note the nonlinearity of both curves. They are nonlinear for the assumption of either constant carbonate-ion concentration or constant total dissolved inorganic carbon concentration. This nonlinear response of pH to changes in atmospheric carbon dioxide partial pressure will be of the greatest importance in the next section, where the validity of either assumption will also be evaluated.

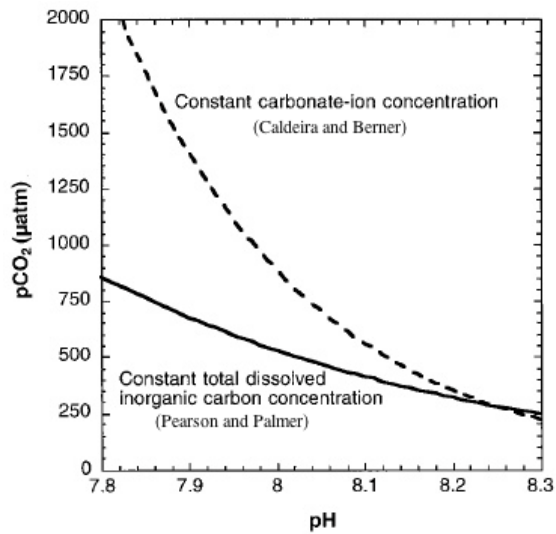

Figure 1. Partial pressure of atmospheric carbon dioxide as a function of pH assuming a pH of 8.25 at a $pCO_2$=280 ppmv for both curves. Adapted from K. Caldeira and R. Berner, *Science* **286**, 2043 (1999).

**Time series data from Ocean Station Aloha**

Time series data from about the mid-1950s to today are available from Ocean Station Aloha in the subtropical North Pacific Ocean. The surface ocean pH data, available from 1990, show an average pH decrease of ~0.03 units. A linear fit made to these data by Feely, et al. [5] resulted in the relation

$$\text{pH(Year date)} = 11.815 - 1.9 \times 10^{-3} \cdot \text{(Year date)}.$$

(5)

For example, pH(2008) = 11.815 − 1.9 × 10⁻³•(2008) = 7.99. The data are shown in Fig. 2. Note that if one uses this linear fit to pH data, the pre-industrial value of pH = 8.25 occurs for the year 1876.

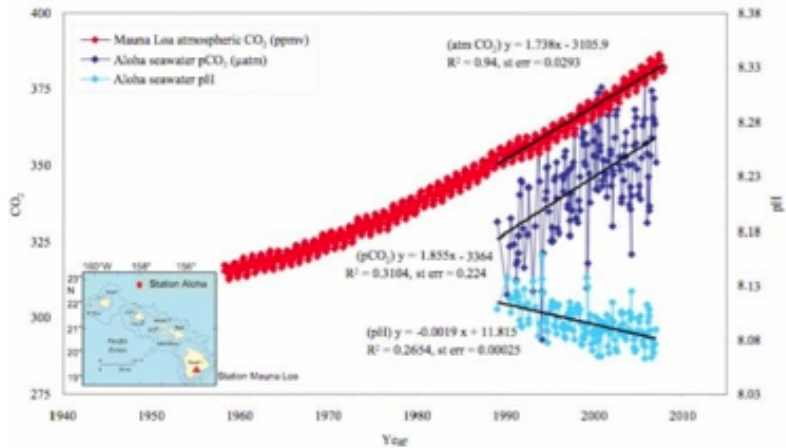

Figure 2. Atmospheric carbon dioxide from Mauna Loa (ppmv) and pCO2 ($\mu$atm), and surface ocean pH time series data from Ocean Station Aloha. From Feely, et al. *PICES Press* **16(1)**, 22-26 (2008).

The Royal Society report [1], referencing Zeebe and Wolf-Gladrow [6], maintains that if there is "no mixing with the deeper oceans . . . it is predicted that pH will fall to below 7.9 by 2100". If one uses Eq. (5) with the date year of 2100, one obtains pH(2100) = 7.8. The Royal Society pH estimate for 2100 is thus consistent with a linear extrapolation of the eighteen years of data from Ocean Station Aloha. Such an extrapolation would appear to be unwarranted or questionable at best. As can be seen from Fig. 1, there is no reason to think the response over the next more than ninety years will be linear.

According to the Ocean Station Aloha data, current pH(2008) ≈ 8.08. For pH = 8.08, the Caldeira and Berner relation shown in Fig. 1 gives a value for $pCO_2$ of 613 ppmv, and that of Pearson and Palmer a value of 413 ppmv. The actual value today is ~384 ppmv. The discrepancy shows that even if the assumption of constant carbonate ion concentration or constant total dissolved inorganic carbon concentration can be justified on theoretical grounds, other factors must be involved.

Caldeira and Wickett [7] project a maximum pH reduction in 2300 of 0.77 units for an atmospheric concentration of carbon dioxide of 1900 ppmv. The latter results from the burning of all fossil-fuel sources estimated at 5,270 GtC. The 0.77 unit reduction in pH could mean either from the pre-industrial value of 8.25 or the current value of 8.08. The former gives a pH of 7.48 and the latter a pH of 7.31. Using Eq. (5) again, one obtains pH(2300) = 7.44. So again, the eighteen years of Ocean Station Aloha or similar data appear to have been linearly extrapolated out to 2300. This is even more questionable than a linear extrapolation to 2100.

The Ocean Station Aloha data imply that the rise of atmospheric carbon dioxide concentration *is* affecting ocean pH. Geographically diverse data would strengthen the case for this argument. However, the current level of understanding of the mixing of surface water with the deep ocean (discussed in the next section), and the relation of atmospheric carbon dioxide concentrations to ocean surface water pH, is not adequate for centennial or longer ocean pH projections. In addition, current extrapolations do not take into account rising water temperature or the response of the biosphere. After all, the biological pump is responsible for some three-quarters of the surface to deep-ocean gradient in dissolved inorganic carbon [8].

**Ocean mixing times**

The Royal Society report assumed no surface water mixing with the deeper oceans. How justified is this assumption? As mentioned earlier, carbon dioxide is generally assumed to mix with the deep ocean on a time scale of about 300 years. This is perhaps long enough to justify not taking ocean mixing into account for projections for the next century or so. But how well is this time scale known?

Deep ocean circulation and mixing is difficult to measure directly, and is generally based on hydrographic measurements such as salinity that give little or no information on rates.

Fortunately, there exist transient tracers such as tritium from nuclear weapons testing in the Pacific Ocean during the late 1950s and early 1960s, peaking in 1963. High tritium readings are now found in the deep Atlantic, and as put by Doney [9], "the deep western boundary current is quite 'leaky': tritium and other tracers introduced into the boundary current are rapidly lost to the interior by turbulent mixing and/or recirculation . . .These results have important implications for understanding the response of the deep North Atlantic to climatic variability on decadal time scales and the invasion of anthropogenic pollutants (e.g. greenhouse $CO_2$) into the deep ocean."

Jenkins and Smethie, Jr. [10] find a 1 Tritium Unit isosurface, below which the seawater has not mixed with tritium and above which it has in the 15 to 20 years between the bomb tests and the survey upon which their article is based. This isosurface lies at depths of 500 to 1,000 meters in the subtropics, but drops to 1,500 to 2,000 meters just south of the Gulf Stream off the New England coast. Near Bermuda, tritium reached the intermediate depths of 1,000-1,500 meters in the late 1970s, and depths of 2,000-2,500 meters in the late 1980s. In the north, the isosurface is at the ocean floor. Jenkins and Smethie, Jr. forcefully state the implications: "The track along which this happens parallels the Gulf Stream Extension North Atlantic Drift. All the waters north of this line have been ventilated to the ocean floor on 10 to 20 year time scales. This is a powerful statement regarding the time scales of ocean ventilation, and has profound implications concerning how rapidly climatic variations can propagate through the oceans".

**Conclusion**

A good deal of research and wide area monitoring is needed before the understanding of the effects of rising carbon dioxide concentration on ocean pH will allow projections adequate to guide public policy decisions. Future research needs and strategies have been detailed in the June 2006 report of an 18-20 April 2005 workshop sponsored by the National Science Foundation, the National Oceanic and Atmospheric Administration, and the U.S. Geological Survey [11].

# APPENDIX

**Seawater Response to Increased Carbon Dioxide**

An increase in atmospheric carbon dioxide will rapidly increase the partial pressure of this gas in the surface waters of the ocean. This in turn will cause surface seawater to become more acidic via the reaction

$$H_2O + (CO_2)_{aq} + CO_3^{2-} \rightarrow 2HCO_3^-.$$

(A1)

This is seen in Figure A1. The relationships shown in Fig. A1 holds for increases of carbon dioxide over a relatively short time.

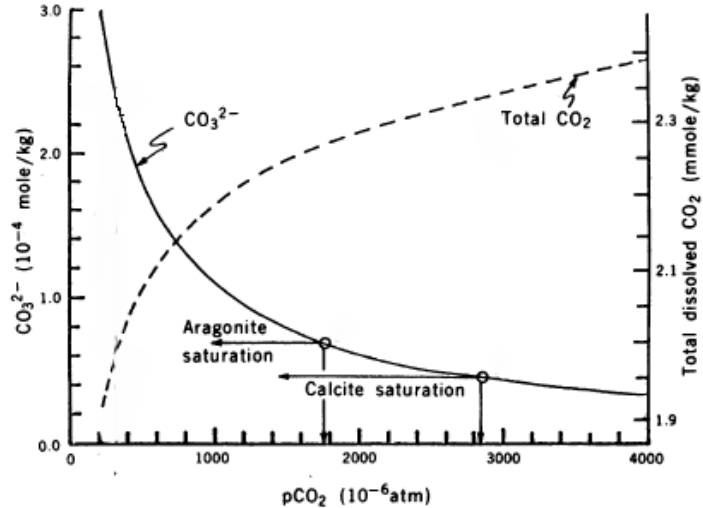

Figure A1. Total dissolved $CO_2$ and $CO_3^{2-}$ concentration as a function of the partial pressure of $CO_2$. Surface sea water at 24 °C, salinity 35 per mil, and an alkalinity of 2.3 milliequivalent per kilogram. Adapted from Broecker, et al. *Science* **206**, 409-418 (1979).

Marine organisms form calcium carbonate in two forms: aragonite and calcite, polymorphs of crystalline $CaCO_3$. Both forms will dissolve unless the surrounding seawater is saturated with respect to the concentration of carbonate ions $CO_3^{2-}$. The aragonite and calcite saturation levels are indicated in Fig. A1.

In the distant past, atmospheric carbon dioxide concentrations were much higher than current values and yet coral reefs thrived. This is despite the fact that ocean pH levels have varied between 7.4 and 8.3 and the lowest values correlate with high carbon dioxide levels. The key to understanding this is that carbon dioxide concentrations changed over periods of time long enough to allow mixing with the deeper parts of the ocean. It is generally believed that this mixing takes about 300 years, while the acidity resulting from dissolved carbon dioxide is partially neutralized by carbonate minerals over a period ~6000 years [12]. Ocean mixing time is a critical parameter for estimating the effect of rising carbon dioxide levels on the ocean, and has been discussed above.

There is some experimental evidence that, rather than pH, the calcification rate of corals responds to the combination of $Ca^{2+}$ and $CO_3^{2-}$ variations. Langdon, et al. [13] have shown that for an experimental coral reef, variations in $Ca^{2+}$ and $CO_3^{2-}$ appear to affect marine calcification rates more or less according to the relation

$$\Omega = \frac{[Ca^{2+}][CO_3^{2-}]}{K_{sp}^*}.$$

(A2)

$\Omega$ is known as the saturation state and $K^*_{sp}$ is the solubility product for either calcite or aragonite. As seawater pH declines as a result rising carbon dioxide concentrations, the concentration of $CO_3^{2-}$ will also fall reducing the calcium carbonate saturation state. Langdon, et al. caution however that the control of calcification in corals and marine calcareous algae by saturation state requires additional verification.

The Royal Society report [1] includes Figure A2 below, showing the relative proportions of $HCO_3^-$, $CO_3^{2-}$, and $CO_2$ dissolved in seawater.

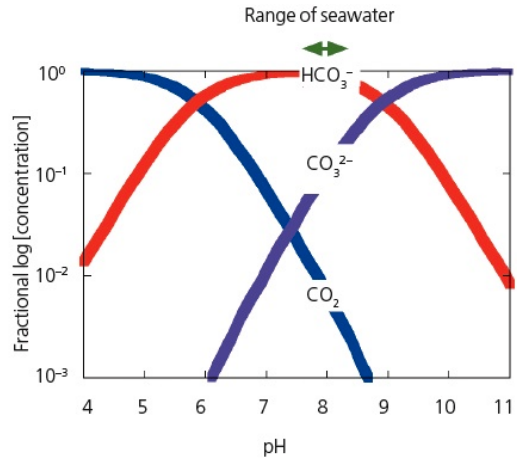

Figure A2. Relative proportions of dissolved inorganic forms of carbon dioxide in seawater. The narrow range 7.5 < pH < 8.5 of pH likely to be found in the oceans at any time is indicated by the arrows at the top of the figure. From the Royal Society report [1].

There are three basic reactions, in addition to Henry's law, that will be needed for what follows. The unconventional symbol ⇔ will be used in equilibrium relations since the usual half arrow forward over a half arrow back is unavailable. The first reaction is then

$$CO_2 + H_2O \Leftrightarrow H_2CO_3.$$

(A3)

Henry's law is

$$pCO_2 = K_H[H_2CO_3],$$

(A4)

where the constant $K_H \sim 29$ (atmos•l/mole). Note that here and in the first dissociation constant for carbonic acid, it is customary for $[H_2CO_3]$ to include dissolved $CO_2$ gas as well as the concentration of $H_2CO_3$.

The second basic reaction along with the first dissociation constant for carbonic acid is

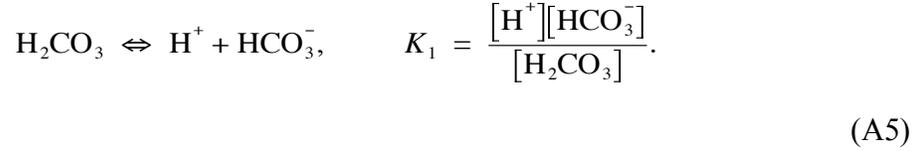

$$H_2CO_3 \Leftrightarrow H^+ + HCO_3^-, \qquad K_1 = \frac{[H^+][HCO_3^-]}{[H_2CO_3]}.$$

(A5)

The third reaction and the second dissociation constant is given by

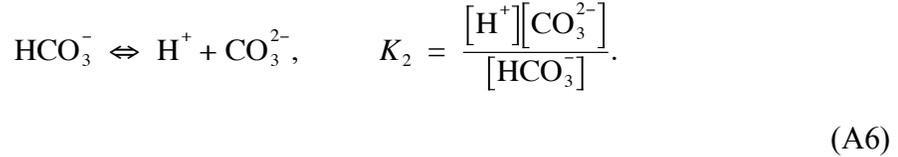

$$HCO_3^- \Leftrightarrow H^+ + CO_3^{2-}, \qquad K_2 = \frac{[H^+][CO_3^{2-}]}{[HCO_3^-]}.$$

(A6)

These reactions show that atmospheric carbon dioxide in its gaseous form is absorbed into seawater according to Henry's law where it reacts with water to form carbonic acid. Some of the carbonic acid dissociates into hydrogen and bicarbonate ions. A portion of the bicarbonate ions dissociate into additional hydrogen and carbonate ions. These reactions are reversible and have associated equilibrium constants. At a pH of ~8.1, about 90% of the carbon is in the form of bicarbonate ions, 9% in the form of carbonate ions, and 1% in the form of dissolved carbon dioxide.

Marine carbonates also react with carbon dioxide through the reaction

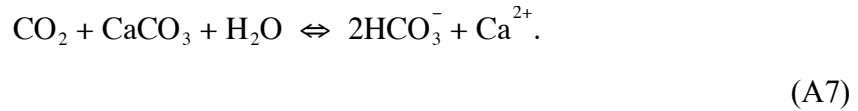

$$CO_2 + CaCO_3 + H_2O \Leftrightarrow 2HCO_3^- + Ca^{2+}.$$

(A7)

This reaction depends on the calcium carbonate saturation state; i.e., the degree of seawater saturation with respect to calcite and aragonite. Since the ratio of calcium to salinity only varies by a few percent, variations in $CO_3^{2-}$ to $K_{sp}*$ are primarily responsible for the saturation state [see Eq. (A2)]. Unless sufficient $CO_3^{2-}$ is present, $CaCO_3$ will dissolve back into the surrounding seawater. This creates a "saturation horizon" for calcite and aragonite, below which $CaCO_3$ cannot form.

**Derivation of the relationship between pH and pCO$_2$**

The relation between pH and pCO$_2$ is crucial for understanding the future impact of rising carbon dioxide concentrations on the ocean. Without this understanding, it is impossible to reliably project seawater pH into the future, particularly beyond a decade or two. Two approaches, which were developed to estimate the pH of Middle Eocene seawater, will be derived here.

The first and second dissociation constants in Eqs. (A5) and (A6) tell us that

$$\frac{K_1}{[H^+]} + \frac{K_1 K_2}{[H^+]^2} = \frac{[CO_3^{2-}] + [HCO_3^-]}{[H_2CO_3]}.$$

(A8)

Adding $1 = \frac{[H_2CO_3]}{[H_2CO_3]}$ to both sides of this equation results in

$$1 + \frac{K_1}{[H^+]} + \frac{K_1 K_2}{[H^+]^2} = \frac{[CO_3^{2-}] + [HCO_3^-] + [H_2CO_3]}{[H_2CO_3]}.$$

(A9)

Now the numerator on the r.h.s. of this equation is the total concentration of dissolved inorganic carbon in seawater. Calling this sum $\Sigma CO_2$, and using Henry's law, results in

$$pCO_2 = K_H[H_2CO_3] = K_H \left(1 + \frac{K_1}{[H^+]} + \frac{K_1 K_2}{[H^+]^2}\right)^{-1} \Sigma CO_2.$$

(A10)

This is the relation given by Pearson and Palmer [3] in their footnote 32 for estimating $pCO_2$ as a function of hydrogen ion concentration and total dissolved carbon. A relation between pH and pCO2 can be obtained by remembering that pH = $-\log[H^+]$, or $[H^+] = 10^{-pH}$. Using this in Eq. (A10) results in

$$pCO_2 = \frac{K_H \Sigma CO_2}{\left(1 + K_1 10^{pH} + K_1 K_2 10^{2pH}\right)}.$$

(A11)

Order of magnitude estimates of the dissociation constants imply that only the second term in the denominator of the r.h.s. is of importance. This results in the relation

$$\text{pCO}_2 = \frac{K_H \Sigma \text{CO}_2}{K_1 10^{\text{pH}}}.$$

(A12)

In their estimates for pCO$_2$ as a function of pH, Pearson and Palmer hold total dissolved carbon constant.

There is an alternative relation that can be derived from Henry's law and Eqs. (A5) and (A6). By simple substitution, one has

$$\text{pCO}_2 = \frac{K_H}{[K_1]}[\text{H}^+][\text{HCO}_3^-] = \frac{K_H}{K_1 K_2}[\text{H}^+]^2[\text{CO}_3^{2-}].$$

(A13)

Caldeira and Berner [4] used this form when objecting to Pearson and Palmer's assumption that total dissolved carbon should be held constant and instead suggested that [CO$_3^{2-}$] should be held constant.

Equation (A13) may also be used to derive a relation between pH and pCO$_2$. Taking the log of both sides of Eq. (A13) and doing a little manipulation results in

$$\text{pCO}_2 = \frac{K_H}{K_1 K_2} \frac{[\text{CO}_3^{2-}]}{100^{\text{pH}}}.$$

(A14)

Equations (A12) and (A14) are the same as Eqs. (2) and (4) in the second section of this paper.

**REFERENCES**

[1] "Ocean acidification due to increasing atmospheric carbon dioxide", Policy document 12/05 (June 2005). ISBN 0 85403 617 2. Available at http://www.royalsoc.ac.uk.

[2] R. E. Zeebe, J. C. Zachos, K. Caldeira, and T. Tyrrell, *Science* **321**, 51-52 (2008).

[3] P. N. Pearson and M. R. Palmer, "Middle Eocene Seawater pH and Atmospheric Carbon Dioxide Concentrations" *Science* **284**, 1824-1826 (1999).

[4] K. Caldeira and R. Berner, "Seawater pH and Atmospheric Carbon Dioxide", *Science* **286**, 2043 (1999).

[5] R. A. Feely, V. J. Fabry, and J. M. Guinotte, *PICES Press* **16**, 22-26 (2008).

[6] R. E. Zeebe and D. Wolf-Gladrow, *CO2 in seawater: equilibrium, kinetics, isotopes, Elsevier Oceanography Series* (Elsevier, Amsterdam 2001).

[7] K. Caldeira and M. E. Wickett, "Anthropogenic carbon and ocean pH", *Nature* **425**, 365 (2003).

[8] J. L. Sarmiento and M. Bender, "Carbon biogeochemistry and climate change", *Photosynthesis Research* **39**, 209-234 (1994).

[9] S. C. Doney, "Bomb Tritium in the Deep North Atlantic", *Oceanography* **5**, 169-170 (1992).

[10] W. J. Jenkins and W. M. Smethie, Jr. "Transient Traers Track Ocean Climate Signals", Oceanus pp. 29-32 (Fall/Winter 1996).

[11] Kleypas, J.A., R.A. Feely, V.J. Fabry, C. Langdon, C.L. Sabine, and L.L. Robbins, 2006. "Impacts of Ocean Acidification on Coral Reefs and Other Marine Calcifiers: A Guide for Future Research", report of a workshop held 18–20 April 2005, St. Petersburg, FL, sponsored by NSF, NOAA, and the U.S. Geological Survey, 88 pp.
Available at: http://www.ucar.edu/communications/Final_**acidification**.pdf

[12] K. Caldeira and G. H. Rau, "Accelerating carbonate dissolution to sequester carbon dioxide in the ocean: Geochemical implications", *Geophys. Res. Lett*. **27**, 225-228 (2000).

[13] C. Langdon, T. Takahashi, C. Sweeney, D. Chipman, and J. Goddard, "Effect of calcium carbonate saturation state on the calcification rate of an experimental coral reef", *Global Biogeochem. Cycles* **14**, 639-654 (2000).